# Observation of the Quantized Hall Insulator in the Quantum Critical Regime of the Two-Dimensional Electron Gas

D.T.N. de Lang, L.A. Ponomarenko, A. de Visser[*]

*Van der Waals-Zeeman Institute, University of Amsterdam*

*Valckenierstraat 65, 1018 XE Amsterdam, The Netherlands*

and

A.M.M. Pruisken

*Institute for Theoretical Physics, University of Amsterdam*

*Valckenierstraat 65, 1018 XE Amsterdam, The Netherlands*

We have investigated the Hall resistance $R_H$ near the plateau-insulator transition of a two-dimensional electron gas in the quantum critical regime. High-field magnetotransport data taken on a low-mobility InGaAs/InP heterostructure with the plateau-insulator transition at a critical field $B_c$ of 17.2 T show that the Hall resistance $R_H$ is quantized at $h/e^2$ near the critical filling fraction ($\nu_c = 0.55$) for $T < 1$ K. By making use of universal scaling functions extracted from the magnetotransport data we show that $R_H$ in the insulating phase in the limit $T \rightarrow 0$ is quantized at $h/e^2$ for all values of the scaling parameter $\Delta\nu/(T/T_0)^\kappa$ with $\Delta\nu = \nu - \nu_c$. However, as a function of $\Delta\nu$ (or magnetic field) the Hall resistance diverges in the limit $T \rightarrow 0$ for all values $\nu < \nu_c$.

---

[*] Corresponding author. E-mail address: devisser@science.uva.nl.

# I. INTRODUCTION

The integral quantum Hall effect observed in a low-mobility two-dimensional electron gas (2DEG) in heterostructures or quantum wells is an outstanding laboratory system to study quantum phase transitions. The plateau-to-plateau (PP) transitions form a series of localization-delocalization transitions at Landau level filling fractions $\nu + 1/2$ ($\nu = 0, 1, 2, ...$) with the magnetic field as tuning parameter [1]. At sufficiently large magnetic fields the series of PP transitions in the two-dimensional electron gas terminates in the plateau-insulator (PI) transition from the $\nu = 1$ plateau to the insulating phase $\nu = 0$. Here the lowest Landau level ($N = 0\uparrow$) is pushed to above the Fermi energy by increasing the magnetic field and the 2DEG becomes an insulator. In recent years, the PI transition has received much attention because of its peculiar transport behavior. Especially, the value of the transverse or Hall resistance $R_{xy}$ at large magnetic fields is much disputed.

In the insulating phase the longitudinal $\sigma_{xx}$ and transverse $\sigma_{xy}$ conductivity both vanish when $T \rightarrow 0$. However, the resistivity tensor $\rho_{ij}$ shows a more intricate behavior. Theoretical models predict that whereas the longitudinal resistivity $\rho_{xx}$ diverges, the Hall resistivity $\rho_{xy}$ remains finite, i.e. close to the classical value $\sim B/ne$ [2] (with $n$ the 2D carrier concentration) or quantized at the $\nu = 1$ plateau value $h/e^2$ [3]. The latter state has been named the "quantized Hall insulator". More recently, a breakdown of the quantized Hall insulator has been anticipated [4,5,6,7]: at large magnetic fields, deep in the insulating phase, the Hall resistivity diverges again when $T \rightarrow 0$.

In the past decade experimental studies of the resistivity tensor near the $\nu = 1$ plateau-to-insulator transition were conducted on various low-mobility semiconductor systems [8,9,10,11,12,13,14,15]. The longitudinal resistivity $\rho_{xx}$ was found to diverge in an exponential fashion, which corroborates a vanishing conductivity tensor. A first signature of $\rho_{xy}$ quantized at $h/e^2$ near the PI transition in a GaAs/AlGaAs heterostructure was reported in Ref.8. Substantial experimental evidence for a quantized Hall insulator was provided by Hilke *et al.* [11], who reported a quantized $\rho_{xy} \sim h/e^2$ over the wide range of filling fractions $0.4 < \nu < 1.5$ in a Ge/SiGe quantum well, with the PI transition at $\nu_c = 0.75$. However, in these studies, the experimental parameters indicate that the two-dimensional electron (GaAs) or hole gas (Ge) is in the semi-classical rather than in the quantum critical regime. Consequently, the observation $\rho_{xy} \sim h/e^2$ may be attributed to transport semi-classical in nature [16]. In this case, the quantized Hall resistance directly follows from the semi-classical universal semicircle relation $\sigma_{xx}^2 +(\sigma_{xy} - e^2/2h)^2 = (e^2/2h)^2$ for the $\nu = 1$ plateau-to-insulator phase transition [11,16,17]. This then prompts the important question of the behavior of the Hall resistivity at the PI transition for a 2DEG in the quantum critical regime.

In this paper we investigate the Hall resistance $R_H$ around the plateau-insulator transition of a 2DEG formed in a low mobility InGaAs/InP heterostructures [18]. We make use of a comprehensive analysis of the magnetotransport data (Refs 19,20) in the framework of the scaling theory of the quantum Hall effect [21], which provides solid evidence that the PI transition occurs in the quantum critical regime. We find that the intrinsic Hall resistance $R_H$ remains quantized at the value $h/e^2$ near the critical filling fraction ($\nu_c = 0.55$) for $T < 0$ and calculate $R_H(B)$ and $R_H(X)$, where $X$ is the scaling parameter $\Delta\nu/(T/T_0)^\kappa$, deep in the insulating phase.

## II. SAMPLE CHOICE AND MAGNETOTRANSPORT EXPERIMENT

The 2DEG studied in the present work was confined at the interface of a low-mobility lattice matched $In_{0.43}Ga_{0.57}As$/InP heterostructure. Since the carriers are located in the $In_{1-x}Ga_xAs$ alloy, they will experience potential scattering of short-ranged random nature. Uncorrelated δ-function-like potential fluctuations are a necessary ingredient for probing quantum critical behavior of the quantum Hall transitions over a relatively wide temperature range, typically up to ~ 4 K [22]. Notice long-ranged potential fluctuations, such as present in remotely doped GaAs/AlGaAs heterostructures, tend to suppress the quantum critical regime to below ~0.2 K [23]. This in turn renders transport experiments at the plateau-insulator phase transition at very low temperatures ($T < 0.1$ K) difficult because of Joule heating in the 2DEG by the excitation current.

The InGaAs/InP heterostructure studied here has an electron density $n = 2.2 \times 10^{11}$ cm$^{-2}$ and a transport mobility $\mu = 16000$ cm$^2$/Vs. It was prepared in the form of a Hall bar with a channel width $W$ of 650 μm and length $L$ between the voltage contacts of ~1100 μm. Previous magnetotransport experiments carried out on the same sample confirmed scaling in the quantum Hall regime [15,18,19,24]. Ref. 24 focused on the PP transitions $\nu = 4\rightarrow3$, $3\rightarrow2$ and $2\rightarrow1$, for which it was demonstrated that the components of the resistivity tensor obey scaling, i.e. the maximum slope in the Hall resistivity $(d\rho_{xy}/dB)_{max}$ and the inverse of the half width of $\rho_{xx}$ between adjacent quantum Hall plateaus $(\Delta B)^{-1}$ both diverge algebraically with temperature as $T^{-\kappa}$ with a critical exponent $\kappa = 0.42$ [22]. In Ref. 15 the magnetic field range was extended up to 20 T such that the PI transition near $\nu_c = 0.5$ ($B_c \sim 17$ T) could be probed as well. The longitudinal resistivity followed an exponential law, from which a critical exponent $\kappa = 0.55$ was extracted, while the transverse resistivity $\rho_{xy}$ (measured for a single magnetic field polarity) was found to diverge for $B > B_c$. In subsequent experiments (partly reported in Ref.18) the transport tensor was measured for both field polarities and it immediately became clear that the diverging $\rho_{xy}$ was due to macroscopic sample inhomogeneities. "Contact misalignment" due to sample inhomogeneities at the Hall contacts (lithographic imperfections of the Hall bar are negligible) is significant and results in mixing a considerable part of $R_{xx}$ into $R_{xy}$. By evaluating the intrinsic Hall resistance $R_H$ in the standard way by averaging over two opposite field polarities, $R_H = [R_{xy}(B\uparrow) + R_{xy}(B\downarrow)]/2$, the Hall resistance was found to be quantized for $T < 1$ K. In the following sections of this paper we investigate the $R_H$ data in detail. However, before doing so we elaborate on the experimental details and expand the experimental results.

The magnetotransport experiments up to $B \sim 20$ T were performed at the High Field Magnet Laboratory of the University of Nijmegen. Low temperatures (0.1-4.2 K) were achieved using a $^3$He-$^4$He refrigerator with a plastic dilution unit in order to reduce the effect of eddy current heating during field sweeps. The four-point resistances of the Hall bar were measured using low-frequency lock-in techniques. By using several lock-in amplifiers at the same time, the longitudinal and transverse voltages could be measured in a single magnetic field sweep. The excitation current (~ 1-5 nA) was determined by measuring the voltage drop over a 10 kΩ resistor placed in series with the sample. Special care was taken to prevent systematic errors in the high-ohmic range due to capacitive coupling of the current and voltage wires to the cryostat ground. This was in part facilitated by the relatively wide channel (650 μm) of the Hall bar, which resulted in a relatively low value $R_{xx} \sim 60$ kΩ at the PI transition. With a low excitation frequency (2.6 Hz) we could measure $R_{xx}$ up to 250-300 kΩ ($\rho_{xx} \leq 5.5$ $h/e^2$). The total resistance of the Hall bar (between the current contacts) then amounts to 1 MΩ.

In Fig.1a-c we show the experimental results for $\rho_{xx} = (W/L)R_{xx}$ and $\rho_{xy} = R_{xy}$ in units of $h/e^2$ in the field range 13-19.5 T, i.e. covering the PI transition. The PP transitions for this sample are characterized by the $\rho_{xx}$ data shown in the left part of Fig.1a up to 9 T. Upon reversal of the magnetic field polarity the $\rho_{xx}$ data for the PI transition turn out to be symmetric, $\rho_{xx}(B\uparrow) = \rho_{xx}(B\downarrow)$, as it should. The critical field $B_c = 17.2$ T is identified by the intersection point of the $\rho_{xx}$ isotherms. At the intersection point the critical resistivity $\rho_{xx,c} \approx h/e^2$ with an experimental uncertainty of 10% (due to the error in the determination of $W/L$). The critical filling factor $\nu_c = 0.55$ is close to the theoretical value of 0.5, which ensures that overlap with (localized states in) the lower Landau level $N= 0\downarrow$ is very small.

Clean Hall data are antisymmetric with respect to the field polarity, $\rho_{xy}(B\uparrow) = -\rho_{xy}(B\downarrow)$. In Fig.1b we have plotted the $|\rho_{xy}|$ data as a function of $|B|$, which shows that a significant symmetric component due to $\rho_{xx}$ is mixed into $\rho_{xy}$. After averaging we obtain the intrinsic Hall resistivity $\rho_H$, shown in Fig.1c (we correct for the effect of a small carrier density gradient in section IV). For $T < 1$ K, $|\rho_{xy}(B\uparrow)|$ and $|\rho_{xy}(B\downarrow)|$ are almost symmetric around the $\nu = 1$ plateau value and consequently $\rho_H \approx h/e^2$ up to 19 T. For $T > 1$ K $|\rho_{xy}(B\uparrow)|$ and $|\rho_{xy}(B\downarrow)|$ are no longer symmetric around the value $h/e^2$ and as a result $\rho_H$ deviates from the quantized value. The higher is the temperature, the larger the deviation. These deviations give us important information, as they yield the "corrections to scaling" as will be discussed in the next section.

**III. SCALING FUNCTIONS**

Let us first recall the principles of scaling. Denoting the longitudinal and Hall resistivities of an ideal homogeneous sample by $\rho_0$ and $\rho_H$, respectively, then at sufficiently low temperatures these quantities with varying $B$ and $T$ become functions of a single scaling variable $X$ only [19]

$$\rho_0(B,T) = \rho_0(X) \quad ; \quad \rho_H(B,T) = \rho_H(X) \qquad (1),$$

where

$$X = \frac{\nu - \nu_c}{\nu_0(T)} \quad ; \quad \nu_0(T) = (T/T_0)^\kappa. \qquad (2).$$

Here $\nu_c \approx 0.5$ for the PI transition and $\kappa$ a universal critical exponent. $T_0$ is a sample dependent temperature scale of the order of the full width of the Landau level. In general, macroscopic sample inhomogeneities, such as contact misalignment and carrier density gradients, result in non-unique values of $\rho_{xx}$ and $\rho_{xy}$ when measured at different contact pairs of the Hall bar [25]. However, under simple circumstances the experimental resistivity tensor is related to the intrinsic resistivities $\rho_0$ and $\rho_H$ by the relation [19]

$$\rho_{ij} = S_{ij}\rho_0(X) + \varepsilon_{ij}\rho_H(X) \qquad (3).$$

Here $\varepsilon_{ij}$ is an antisymmetric tensor and $S_{ij}$ the "stretch tensor", which describe the sample imperfections. Recently, we have developed [19,20] a method to extract $\rho_0$ and $\rho_H$ from the measured resistivity tensor $\rho_{ij}$ for the PI transition. This procedure largely relies on a fundamental symmetry in the quantum Hall problem namely "particle-hole" symmetry, $\sigma_0(X) = \sigma_0(-X)$ and $\sigma_H(X) = 1-\sigma_H(-X)$. For the PI transition "particle-hole" symmetry translates to $\rho_0(X) = 1/\rho_0(X)$ and a quantized value $\rho_H = 1$ (from now on we work in units $h/e^2$). Notice that for the PP transitions the steps in $\rho_H$ complicate the problem considerably [26].

The scaling results obtained in Ref.[19,20] are:

$$\rho_0(X) = e^{-X-\gamma X^3 - O(X^5)} \qquad (4),$$

$$\rho_H(X,\eta) = 1 + \eta(T)\rho_0(X) \quad ; \quad \eta(T) = \left(\frac{T}{T_1}\right)^{y_\sigma} \qquad (5).$$

Eq.4 describes the exponential dependence of $\rho_0$ for the various isotherms. By plotting $\ln\rho_0$ versus $X$, $1/\nu_0(T)$ is determined. Subsequently, the critical exponent $\kappa$ and the temperature $T_0$ are extracted by plotting $1/\nu_0(T)$ versus $T$ on a log-log scale. The third order term $-\gamma X^3$ in the exponential is a small correction term in the regime of interest. The amplitude of Eq.4 is given by $\rho_0(B_c) = 1$ (in units of $h/e^2$) in agreement with the experimental data (see Fig.1).

Eq.5 describes the corrections to quantization of the Hall resistivity $\rho_H = 1$ at higher $T$ (see Fig.1). The correction term $\eta(T)\rho_0(X)$ indicates that under "ordinary" quantum Hall conditions ($X \neq 0$) the corrections are exponential in $T$, while at the critical point ($X = 0$) they are algebraic in $T$. The function $\eta(T)$ can be determined by factoring out $\rho_0(X)$ from Eq.5. In the presence of a small carrier density gradient as observed in our sample, $\rho_0(X)$ should be replaced by $\rho_0(X')$, where $X'$ represents a local value, which differs slightly from $X$ [19]. $X'$ is the same as in Eq.2, but with $v_c'=v_c/(1+\varepsilon_x)$ and $T_0'=T_0(1+\varepsilon_x)^\kappa$. For our InGaAs/InP structure $\varepsilon_x \approx -0.03$ [19,20]. Factoring out $\rho_0(X')$ leads to a collapse of *all* the $\rho_H$ data in the vicinity of the critical point onto a single curve $\eta(T) = (1-\rho_H)/\rho_0(X') = (T/T_1)^{y_\sigma}$ with critical exponent $y_\sigma$ and a temperature scale $T_1$. $T_1$ indicates the relevant $T$ range for scaling. The collapse of the data in the range $-0.025 < \Delta v < 0.05$ is shown in Fig.1d. Note that the values of the experimental Hall data $\rho_H(X')$, shown in Fig.1c, are slightly smaller than the intrinsic values $\rho_H(X)$, as $\rho_0(X') < \rho_0(X)$. Having extracted $\eta(T)$ and $\rho_0(X)$ from the experimental data we now have a precise description of the *intrinsic* Hall resistivity $\rho_H$, that will be explored in the next section.

## IV. THE HALL RESISTANCE IN THE ASYMPTOTIC LIMIT

The experimental data reported in Fig.1 show that the Hall resistivity $\rho_H \approx 1$ in the vicinity of the PI transition at the lowest temperatures ($T < 1$ K). Deviations from quantization become significant at $T > 1$ K. Because the resistance of the sample increases exponentially with magnetic field, the filling fraction range over which the insulating phase is probed is limited ($v > 0.50$). However, since we know the exact mathematical expression for $\rho_H$, Eq.5, predictions for $\rho_H$ in the insulating phase can be made. In the following we make use of the scaling functions Eqs 1,2,4,5 with critical exponents $\kappa = 0.57 \pm 0.02$ and $y_\sigma = 2.4 \pm 0.1$ and characteristic temperatures $T_0 = 188 \pm 20$ K and $T_1 = 9.2 \pm 0.3$ K reported for our sample in Refs 19 and 20. In the regime of interest here $|X| < 5$ and we neglect the small third order correction term ($\gamma = 0.002 \pm 0.001$) in the exponential of Eq.4. We evaluate the Hall resistance for three different experimental situations, and investigate the dependence of $\rho_H$ as a function of the scaling parameter $X$ and as a function of $B$ (or rather $v$).

### A. Critical point

At the critical point $X = 0$ ($\Delta v \equiv v - v_c = 0$) and Eq.5 reduces to

$$\rho_H = 1 + \left(\frac{T}{T_1}\right)^{y_\sigma} \qquad (6).$$

Hence, we obtain $\rho_H = 1$ for $T \to 0$.

**B. Quantum Hall regime**

In the quantum Hall regime $X > 0$ ($\Delta \nu > 0$). We rewrite Eq.5 as

$$\rho_H = 1 + e^{-\frac{\Delta \nu}{\nu_0(T)} - \gamma \left(\frac{\Delta \nu}{\nu_0(T)}\right)^3 + y_\sigma \ln(T/T_1)} \qquad (7).$$

With $\gamma = 0.002$, $y_\sigma = 2.4$ and $T_1 = 9.2$ K the three terms in the exponential are negative and $\rho_H = 1$ for $T \to 0$.

**C. Insulator regime**

In the insulator regime $X < 0$ ($\Delta \nu < 0$). Eq.7 now tells us that $\rho_H$ diverges for large values of $X$. However, an important observation is made for $\rho_H(X)$ in the limit $T \to 0$. In Fig.2 we have plotted $\rho_H(X)$ computed with help of Eqs 4,5 with the parameters listed above at five different temperatures in the range 0.8-4.2 K. For simplicity we here take $\gamma = 0$. The scaling behavior shows that for every fixed $X$ $\rho_H(X) \to 1$ in the limit $T \to 0$. Thus $\rho_H(X)$ is quantized in the insulating phase in the asymptotic limit $T \to 0$. In the upper panel of Fig.2 we have plotted the universal scaling curve $\rho_0(X)$ obtained with $\kappa = 0.57$ and $T_0 = 188$ K. In the quantum critical regime the $\rho_0$ isotherms in the temperature interval 0.26-4.2 K collapse to the universal curve Eq.4, while the $\rho_H(X)$ curves represent the corrections to scaling (Eq.5) and "collapse" in the limit $T \to 0$ to $\rho_H(X) = 1$.

A crucial difference is observed when we consider $\rho_H$ as a function of $\Delta \nu$ (or magnetic field). This is demonstrated in Fig.3, where we have plotted the calculated values of $\rho_H(\Delta \nu)$. The $\rho_H(\Delta \nu)$ isotherms intersect, which implies that $\rho_H$ is not a monotonous function of $T$. Upon lowering $T$ $\rho_H$ decreases, but below a certain threshold $T$ $\rho_H(\Delta \nu)$ increases again (see also the next section). Thus in the insulating phase $\rho_H(\Delta \nu)$ diverges in the limit $T \to 0$.

This leads to the following important statement: the Hall resistance for $T\to 0$ in the insulating phase is quantized as a function of the scaling parameter $\Delta\nu/\nu_0$, but diverges as a function of the magnetic field.

**V. THE HALL RESISTANCE AS A FUNCTION OF TEMPERATURE**

The analytical expression for $\rho_H$ Eq.5 may also serve to calculate the temperature variation of $\rho_H$ at fixed values of the magnetic field. In Fig.4 we show $\rho_H$ up to ~ 10 K (i.e. up to approximately $T_1$) for five different filling fractions in the magnetic field range 11.7-32.5 T. The solid line at $\nu_c$ separates the region $\Delta\nu > 0$, where $\rho_H$ approaches 1 (in units of $h/e^2$) for $T\to 0$, from the region $\Delta\nu < 0$ in which $\rho_H$ becomes infinite for $T\to 0$. For negative $\Delta\nu$ $\rho_H(T)$ shows a pronounced minimum at $T_{min} \approx T_0 (-\kappa\Delta\nu/y_\sigma)^{1/\kappa}$. Below $T_{min}$ the 2DEG behaves as a "classical" insulator and above $T_{min}$ as a Hall insulator.

**VI. DISCUSSION AND SUMMARY**

Magnetotransport experiments on our quantum critical InGaAs/InP heterostructure provide solid evidence for a quantized Hall resistance in the temperature range 0.1-1 K in the insulating state (albeit only down to $\nu$ ~ 0.5). In the quantum critical regime of the plateau-insulator transition the relevant variable is $X$ rather than $B$. By making use of universal scaling functions we are able to predict the behavior relatively far away ($|X| < 5$) from the critical point. We find that at fixed $X$ $\rho_H\to 1$ when $T\to 0$. Notice, when keeping $X$ fixed also $\Delta\nu \to 0$ when $T\to 0$. The function $\rho_H(X)$ in fact tells us how the critical point is approached. On the other hand, in magnetotransport experiments one measures $\rho_H(B)$, which in the limit $T\to 0$ diverges above ($\nu < \nu_c$) and is quantized below ($\nu > \nu_c$) the PI transition (see Fig.4).

The experimental parameters for this sample are favorable for studying the PI transition. The critical resistance $R_{xx}$~ 60 k$\Omega$ at $B_c$ is low, which enables us to acquire ac-resistance data over a significant field range in the insulating phase. Also, contact misalignment is relatively small (the field symmetric part in $\rho_{xy}$ is about 10% of $\rho_{xx}$) which permits the extraction of proper $\rho_H$ data by averaging $R_{xy}(B\uparrow)$ and $R_{xy}(B\downarrow)$.

Recently, we have investigated the Hall resistance near the PI transition in a second low-mobility InGaAs/InP heterostructures. This sample with electron density $n = 3.4\times 10^{11}$ cm$^{-2}$ and transport mobility $\mu$ = 34000 cm$^2$/Vs was previously studied by Wei *et al.* in their pioneering experiments on scaling and universality of the PP transitions [22]. Magnetotransport experiments up to 30 T in the temperature interval 0.14-4.2 K located the PI transition at $B_c$ = 26.4 T [27], i.e. at a critical filling fraction $\nu_c$ ~ 0.53. The analysis of the transport data within the scaling scenario revealed $\kappa$ = 0.58±0.02 for the PI transition. The Hall resistance, extracted at a few temperatures $T$ < 1 K, was found to be approximately quantized up to $B$ = 30 T ($\nu_c$ ~ 0.47): upon approaching the PI transition $\rho_H$ starts to deviate slightly from the value $h/e^2$, with a maximum *negative* deviation of ~ 2% at 30 T. In this case the small negative deviation is probably due to the relatively short distance between the current and voltage contacts of the Hall bar, as was indicated by numerical simulations of the quantum Hall transport problem [27]. Unfortunately for this sample contact misalignment was quite large and we could not determine the corrections to scaling properly.

We have also studied the PI transition in a low-mobility InGaAs/GaAs quantum well for which the carrier density could be controlled by persistent photoconductivity [28]. The longitudinal and Hall resistances were measured in detail for a sample with $n$ = 2.0x10$^{11}$ cm$^{-2}$ and $\mu$ = 16000 cm$^2$/Vs and the PI transition at $B_c$ = 15.7 T ($\nu_c$ = 0.53 T). We extracted $\kappa$ = 0.58±0.02 from the scaling analysis of $\rho_{xx}$ for the PI transition. At the lowest temperatures $T$ ≤ 0.2 K the Hall resistance was found to be quantized up to $\nu$ = 0.46. From the deviations of quantization at higher temperatures the critical exponent $y_\sigma$ was estimated at 2.6±0.5 and $T_1$ = 4.5±0.8 K [27,28]. The values for the critical exponents $\kappa$ and $y_\sigma$ are very similar to those reported for the InGaAs/InP heterostructure [19].

Having extracted universal quantum critical behavior from magnetotransport data in several low-mobility semiconductor structures, we conclude that the concurrent quantization of $\rho_H(X)$ in the vicinity of the PI transition and $T \to 0$ is a robust phenomenon. Also the diverging Hall resistance $\rho_H(B)$ in the insulating phase is a solid result. We have checked that small changes (of the order of the experimental errors) in the parameters that enter Eq.7 do not affect the basic dependence of $\rho_H$ on $X$ or $\Delta\nu$ shown in Figs.2 and 3. This is also true if we use a much smaller value of $\kappa$, e.g. $\kappa$ = 0.42 previously reported for the PP transitions of these InGaAs/InP heterostructures [22,24].

We stress that our magnetotransport experiments carried out on different samples [18,19,20,15,27,28] provide solid evidence for an intrinsic critical exponent $\kappa$ = 0.58 for the PI transition. The critical exponent is obtained directly from the $\rho_{xx} \approx \rho_0(\nu)$ data. By analyzing the magnetotransport problem in Hall bar geometry taking into account small macroscopic sample inhomogeneities, such as density gradients and contact misalignment, we have recently shown [19,20,26] that the critical exponent for the PI transition is robust against sample imperfections, while the exponent for the PP transitions is not. This is corroborated by numerical simulations of the quantum Hall transport problem in Hall bar geometry [27]. For instance, by introducing a carrier density gradient along the Hall bar the $\kappa$-value for the 2→1 PP transition becomes smaller. The reduction of $\kappa$ is typically of the order of 10% for a density gradient along the Hall bar of 2% as present in our samples. Thus macroscopic sample inhomogeneities may be at the origin of the smaller $\kappa$-values for the PP transitions, compared to the intrinsic value $\kappa = 0.58$ for the PI transition. Notice in Ref. 15 the reverse incorrect conclusion was drawn, because at that time experiments were done for one polarity of the magnetic field only. We conclude that, because of sample inhomogeneities, the current experiments do not allow to make a decisive conclusion as regards the universality of PP and PI transitions, predicted by the scaling theory of the quantum Hall effect.

For PP transitions a wide range of $\kappa$-values ($0.3 < \kappa < 0.8$) has been reported in the literature (see e.g. Ref. 29). However, many of these experiments were carried out on GaAs/AlGaAs heterostructures where long-ranged potential fluctuations due to remote doping, in addition to macroscopic sample inhomogeneities, complicate the observability of scaling due to a cross-over from mean-field behavior at high $T$ to critical behavior at low $T$ [30]. Clearly, more experimental work is needed to settle the issue of the precise value of $\kappa$ and its universality. A significant step forward in this respect was recently made by Li *et al*. [31] who carried out state of the art experiments on $Al_xGa_{1-x}As/Al_{0.33}Ga_{0.67}As$ quantum wells with controlled short-ranged random-alloy potential fluctuations. Their magnetotransport data show universal scaling, i.e. for several PP transitions in higher Landau levels ($N = 1\downarrow$ and higher), with $\kappa = 0.42$, for an aluminum contents in the range $0.0065 < x < 0.016$. For larger values of $x$, $\kappa$ increases to ~0.58, which was attributed to the break-down of universal scaling due to clustering of Al atoms. However, transitions in lower Landau levels (e.g. the PI transition) were not studied in these samples because of the appearance of fractional quantum Hall states.

The theoretical studies [7] to model the quantum Hall insulator are based on transport in random networks, which consist of weakly coupled puddles of the quantum Hall liquid. In the absence of quantum interference effects the Hall resistance is predicted to be quantized in the quantum Hall and insulator phase. However, in the quantum coherent regime quantum interference destroys the quantized Hall insulator as a consequence of localization and the Hall resistance diverges [4]. Whether transport is coherent or not depends on the ratio of the puddle size $L_p$, which is a measure for the scale of the potential fluctuations, and the dephasing length $L_\varphi$. At sufficiently low temperatures $L_\varphi > L_p$ (since $L_\varphi \to \infty$ for $T \to 0$) and the Hall resistance diverges. This is at variance with our scaling result $\rho_H$ quantized. Also far above the transition $L_\varphi > L_p$ (because of the small puddle size $L_p$) and the Hall resistance diverges [4]. For this regime it is predicted that $\rho_H \sim \rho_0^\gamma$ with $\gamma$ ranging from 0.26-0.5 [4,5,6]. This relation differs from our scaling result $\rho_H \sim 1+\eta(T)\rho_0$ which holds in the vicinity of the PI transition. Obviously, we did not measure the magnetotransport properties to deep in the insulating phase. However, a diverging Hall resistance far above the transition is consistent with the extrapolation of our $\rho_H(\Delta \nu)$ curves.

In conclusion, we have investigated the Hall resistance $R_H$ near the plateau-insulator transition at $B_c = 17.2$ T in a low-mobility InGaAs/InP heterostructure. Our magnetotransport experiments probe the two-dimensional electron gas under quantum critical conditions and contrast previous work in which transport is semi-classical in nature. The data show that the Hall resistance is quantized at the value $h/e^2$ near the critical filling factor ($\nu_c = 0.55$) at the lowest temperatures measured 0.1-1 K. By extrapolating the universal scaling function for $R_H$ extracted from the magnetotransport data [19], we find that $R_H$ as a function of the scaling parameter $X = \Delta \nu / \nu_0$ is quantized at the value $h/e^2$ for all values $X \leq 0$. However, $R_H$ diverges as a function of the filling factor or magnetic field. In other words, in the insulating phase the Hall resistance in the limit $T \to 0$ is always quantized as a function of $X$ but not as a function of magnetic field. Within the scaling theory of the quantum Hall effect $R_H(X)$ is the proper function to explore.

**Acknowledgements:** This work was part of the research program of FOM (Dutch Foundation for Fundamental Research of Matter). C. Possanzini and S.M. Olsthoorn are gratefully acknowledged for their help in carrying out the experiments at the High Field Magnet Laboratory of the Radboud University Nijmegen.

**REFERENCES**

[1]   A.M.M. Pruisken, in: *The Quantum Hall Effect*, Eds R.E. Prange and S.M. Girvin (Springer Verlag, Berlin 1990), pp.117-173.

[2]   S. Kivelson, D.H. Lee and S.C. Zhang, Phys. Rev. B 46, 2223 (1992).

[3]   E. Shimshoni and A. Auerbach, Phys. Rev. B 55, 9817 (1997).

[4]   L.P. Pryadko and A. Auerbach, Phys. Rev. Lett. 82, 1253 (1999).

[5]   U. Zülicke and E. Shimshoni, Phys. Rev. B 63, 241301 (2001).

[6]   P. Cain and R.A. Römer, Europhys. Lett. 66, 104 (2004).

[7]   For a recent review see: E. Shimshoni, Mod. Phys. Lett. B 18, 923 (2004).

[8]   D. Shahar, D.C. Tsui, M. Shayegan, J.E. Cunningham, E. Shimshoni and S.L. Sondhi, Solid State Comm. 102, 817 (1997).

[9]   W. Pan, D. Shahar, D.C. Tsui, H.P. Wei and M. Razeghi, Phys. Rev. B 55, 15431 (1997).

[10]  M. Hilke, D. Shahar, S.H. Song, D.C. Tsui, Y.H. Xie and D. Monroe, Phys. Rev. B 56, R15545 (1997).

[11]  M. Hilke, D. Shahar, S.H. Song, D.C. Tsui, Y.H. Xie and D. Monroe, Nature 395, 675 (1998).

[12]  R.B. Dunford, N. Griffin, M. Pepper, P.J. Phillips and T.E. Whall, Physica E 6, 297 (2000).

[13]  S.Q. Murphy, J.L. Hicks, W.K. Liu, S.J. Chung, K.J. Goldammer and M.B. Santos, Physica E 6, 293 (2000).

[14]  C.F. Huang, Y.H. Chang, H.H. Cheng, C.T. Liang and G.J. Hwang, Physica E 22, 232 (2004).

[15]  R.T.F. van Schaijk, A. de Visser, S.M. Olsthoorn, H.P. Wei and A.M.M. Pruisken, Phys. Rev. Lett. 84, 1567 (2000).

[16]  A. M. Dykhne and I. M. Ruzin, Phys. Rev. B 50, 2369 (1994).

[17]  D. Shahar, D.C. Tsui, M. Shayegan, E. Shimshoni and S.L. Sondhi, Phys. Rev. Lett. 79, 479 (1997).

[18]  D.T.N. de Lang, L. Ponomarenko, A. de Visser, C. Possanzini, S.M. Olsthoorn and A.M.M. Pruisken, Physica E 12, 666 (2002).

[19] A.M.M. Pruisken, D.T.N. de Lang, L.A. Ponomarenko and A. de Visser, Solid State Comm. 137, 540 (2006).

[20] D.T.N. de Lang, Ph.D. Thesis, University of Amsterdam (2005), unpublished.

[21] A.M.M. Pruisken, Phys. Rev. Lett. 61, 1297 (1988).

[22] H.P. Wei, D.C. Tsui, M.A. Paalanen and A.M.M. Pruisken, Phys. Rev. Lett. 61, 1294 (1988).

[23] H.P. Wei, S.Y. Lin, D.C. Tsui and A.M.M. Pruisken, Phys. Rev. B 45, R3926 (1992).

[24] S.W. Hwang, H.P. Wei, L.W. Engel, D.C. Tsui and A.M.M. Pruisken, Phys. Rev. B 48, 11416 (1993).

[25] L.A. Ponomarenko, D.T.N. de Lang, A. de Visser, V.A. Kulbachinskii, G.B. Galiev, H. Künzel and A.M.M. Pruisken, Solid State Comm. 130, 705 (2004).

[26] B. Karmakar, M.R. Gokhale, A.P. Shah, B.M. Arora, D.T.N. de Lang, L.A. Ponomarenko, A. de Visser and A.M.M. Pruisken, Physica E 24, 187 (210).

[27] L.A. Ponomarenko, Ph.D. Thesis, University of Amsterdam (2005), unpublished.

[28] L.A. Ponomarenko, D.T.N. de Lang, A. de Visser, D. Maude, B.N. Zvonkov, R.A. Lunin and A.M.M. Pruisken, Physica E 22, 326 (2004).

[29] S. Koch, R.J. Haug, K. von Klitzing and K. Ploog, Phys. Rev. B 43, 6828 (1991).

[30] A.M.M. Pruisken, B. Skoric and M.A. Baranov, Phys. Rev. B 60, 16838 (1999).

[31] W. Li, G.A. Csáthy, D.C. Tsui, L.N. Pfeiffer and K.W. West, Phys. Rev. Lett. 94, 206807 (2005).

**FIGURE CAPTIONS**

Figure 1  (a) Longitudinal resistivity as a function of magnetic field for the InGaAs/InP heterostructure at the plateau-plateau transitions ($B < 9$ T) and the plateau-insulator transition ($B > 13$ T) [18]. The curves labeled a,b,…f correspond to $T$ = 0.38, 0.65, 1.1, 2.1, 2.9 and 4.2 K, respectively. The PI transition is located at $B_c = 17.2$ T.

(b) The absolute value of the transverse resistivity as a function of the magnetic field for two opposite polarities $B\uparrow$ ($B_{up}$) and $B\downarrow$ ($B_{down}$).

(c) The Hall resistivity obtained by after averaging $\rho_{xy}$ over the opposite field polarities.

(d) Collapse (closed squares) of the $\rho_H$ data in the interval $-0.025 < \Delta\nu < 0.05$ at fixed $T$, onto a single curve $\eta(T)$ (solid line). See text.

Figure 2  Lower panel: Hall resistivity $\rho_H$ plotted as a function of the scaling variable $X$. Symbols: experimental data measured at $T$ = 0.8, 1.1, 2.1, 2.9, 4.2 K. Solid lines: $\rho_H$ calculated using Eq.7 with parameters as indicated in the figure and $\gamma = 0$. In the limit $T\rightarrow 0$ $\rho_H \rightarrow 1$ for every value of $X$.

Upper panel: Universal scaling curve $\rho_0(X)$ obtained after a collapse of all $\rho_0$ isotherms in the $T$-range 0.26-4.2 K using $\kappa = 0.57$ and $T_0 = 188$ K.

Figure 3  Hall resistivity $\rho_H$ plotted as a function of $\Delta\nu$. Symbols: experimental data measured at $T$ = 0.8, 1.1, 2.1, 2.9, 4.2 K. Solid lines: $\rho_H$ calculated using Eq.7 with parameters as indicated and $\gamma = 0$. In the insulating phase ($\nu < \nu_c$) $\rho_H$ diverges in the limit $T\rightarrow 0$.

Figure 4  Calculated temperature variation (up to $\sim T_1$) of $\rho_H$ for $\Delta\nu$ = -0.25, -0.15, 0, 0.15 and 0.25 (corresponding to $B$ = 32.5, 23.9, 17.2, 13.4 and 11.7 T, respectively) with parameters given in Section IV. For $\nu < \nu_c$ $\rho_H \rightarrow \infty$ when $T\rightarrow 0$, while for $\nu \geq \nu_c$ $\rho_H \rightarrow 1$ when $T\rightarrow 0$.

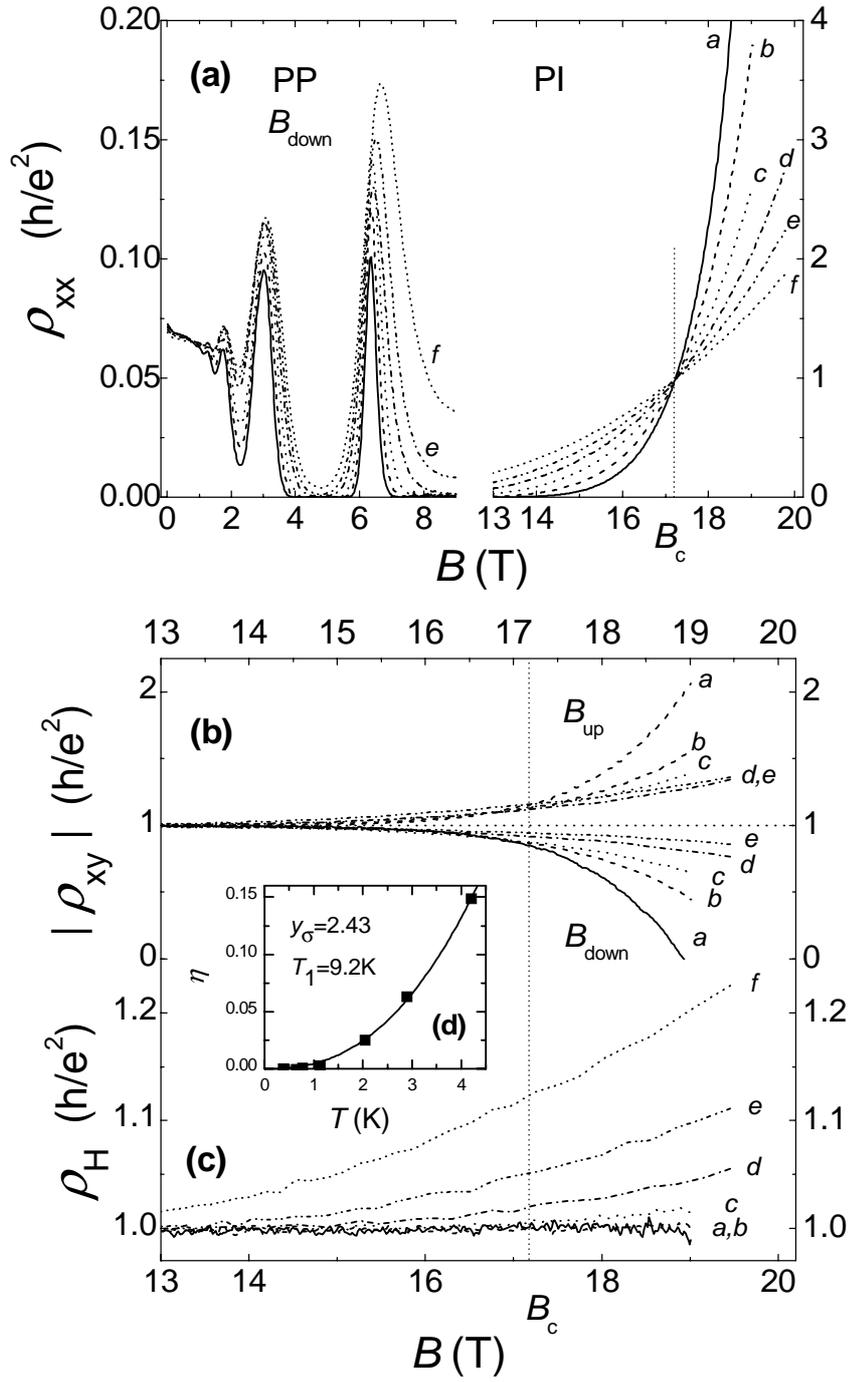

**Fig.1**

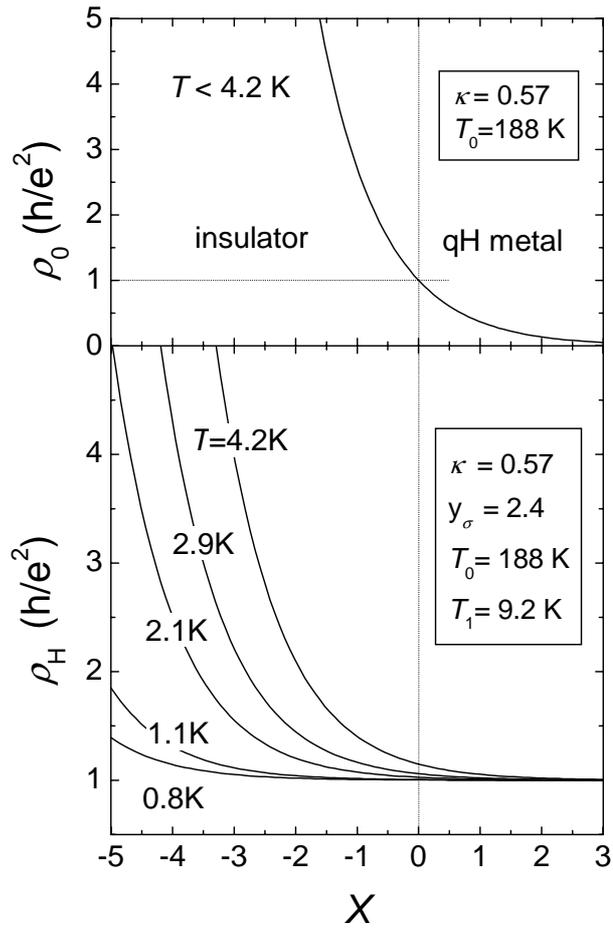

**Fig.2**

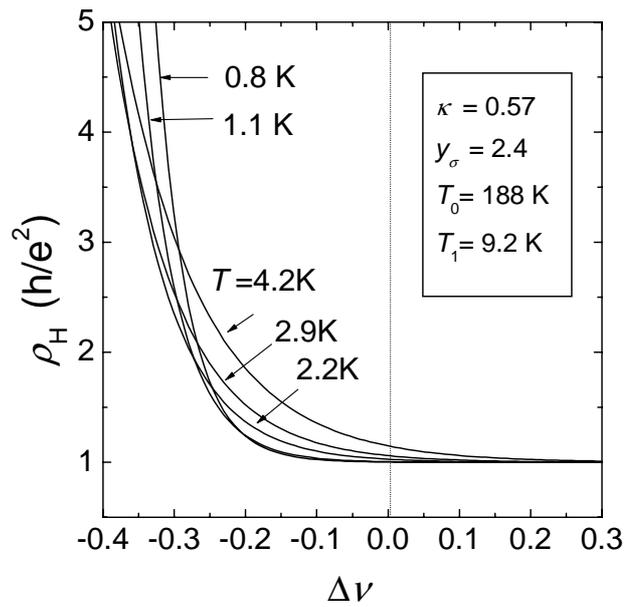

**Fig.3**

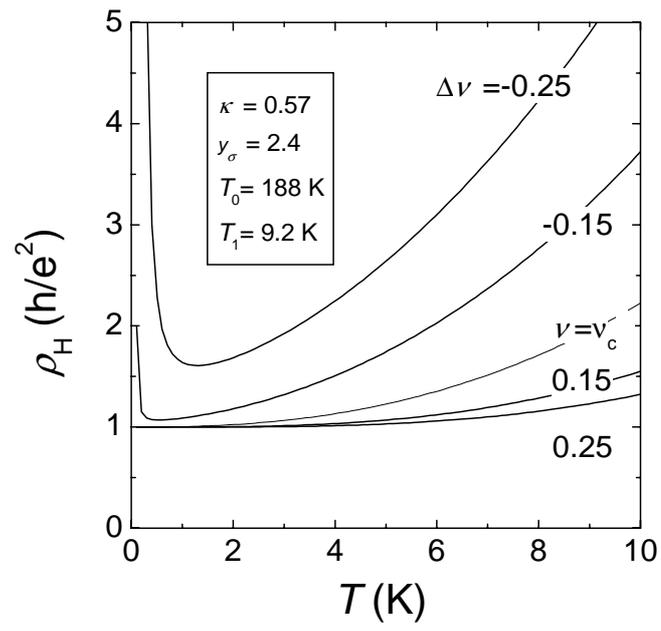

**Fig.4**